\documentclass[conference,a4paper]{IEEEtran}

\usepackage[utf8]{inputenc} 
\usepackage[T1]{fontenc}
\usepackage{color}
\usepackage[dvipsnames]{xcolor}
\usepackage{url}              % provides \url{...}
\usepackage{cite}             % improves presentation of citations

\usepackage[cmex10]{amsmath}  % Use the [cmex10] option to ensure complicance
\usepackage{amsthm}                              % with IEEEXplore (see bare_conf.tex)
\usepackage{mleftright}       % fix to wrong spacing of \left-,
\mleftright                   % \middle- \right-commands 

\usepackage{graphicx}         % provides \includegraphics{...} to
\usepackage{hyperref}
\hypersetup{colorlinks,
   colorlinks,
    citecolor=red,
    filecolor=green,
  linkcolor=blue,
    linktoc=page,
   urlcolor=blue
}                              % include graphics (pdf format)
\usepackage{booktabs}         % fixes poor spacing in tables and
\newtheorem{proposition}{{Proposition}}

\newtheorem{theorem}{{Theorem}}
\newtheorem{lemma}{{Lemma}}

\newtheorem{corollary}{{Corollary}}
\newtheorem{remark}{{Remark}}

\usepackage{physics}
\renewcommand\dv[3][]{\derivative[#1]{#2}{#3}{}}
\usepackage{mathtools}

\DeclareMathOperator{\IB}{IB}
\newcommand{\XX}{\mathcal{X}}
\newcommand{\YY}{\mathcal{Y}}
\newcommand{\TT}{\mathcal{T}}
\makeatletter
\newcommand{\oset}[2]{%
  {\mathop{#2}\limits^{\vbox to -1\ex@{\kern-\tw@\ex@
   \hbox{\scriptsize #1}\vss}}}}
\makeatother
\newcommand{\conphi}{\oset{$\frown$}{\varphi}}

\begin{document}

\title{The Cardinality Bound on the Information Bottleneck Representations is Tight}

%%%%%%
\IEEEoverridecommandlockouts
\author{
\IEEEauthorblockN{Etam Benger}
\IEEEauthorblockA{\textit{CSE,} %\\
\textit{The Hebrew University}\\
Jerusalem, Israel \\
etam.benger@mail.huji.ac.il}
\and
\IEEEauthorblockN{Shahab Asoodeh}
\IEEEauthorblockA{\textit{CS,} %\\
\textit{McMaster University}\\
Hamilton, ON, Canada \\
asoodehs@mcmaster.ca}
\and
\IEEEauthorblockN{Jun Chen}
\IEEEauthorblockA{\textit{ECE,} %\\
\textit{McMaster University}\\
Hamilton, ON, Canada \\
chenjun@mcmaster.ca}
\thanks{The work of E.B.\ was supported in part by the Pazy Foundation, a gift to the McCourt School of Public Policy and Georgetown University, Simons Foundation Collaboration 733792, Israel Science Foundation (ISF) grant 2861/20, a grant from the Israeli Council for Higher Education, the Center for Interdisciplinary Data Science Research (CIDR) at the Hebrew University, and a Google research grant. The work of S.A.\ was supported in part by NSERC Canada.}
}
\maketitle

%%%%%
\begin{abstract}
  The information bottleneck (IB) method aims to find compressed representations of a variable $X$ that retain the most relevant information about a target variable $Y$. We show that for a wide family of distributions -- namely, when $Y$ is generated by $X$ through a Hamming channel, under mild conditions -- the optimal IB representations require an alphabet strictly larger than that of $X$. This implies that, despite several recent works, the cardinality bound first identified by Witsenhausen and Wyner in 1975 is tight. At the core of our finding is the observation that the IB function in this setting is not strictly concave, similar to the deterministic case, even though the joint distribution of $X$ and $Y$ is of full support. Finally, we provide a complete characterization of the IB function, as well as of the optimal representations for the Hamming case.
\end{abstract}

\section{Introduction}

Let $X$ be a random variable defined over a finite alphabet $\XX$ with marginal distribution $p_X$, and let $Y$ be a random variable generated by $X$ through a channel $p_{Y|X}:\XX\rightarrow\YY$. The information bottleneck (IB) function, introduced by \cite{tishby99ib}, is defined as
\begin{equation}\label{eq:IB}
    \IB(R) = \IB(R, p_X, p_{Y|X}) = \max_{\substack{p_{T|X}:\\
    Y-X-T,\\
    I(X;T) \leq R}} I(Y;T) ,
\end{equation}
for $R\in[0, H(X)]$, where $Y-X-T$ means that $Y$, $X$, and $T$ form a Markov chain in that order.
We refer to the variable $T$ as a representation of $X$. This problem is usually solved by minimizing the Lagrangian
\begin{equation}
    \mathcal L[p_{T|X}, \beta] = I(X;T) - \beta I(Y;T)
\end{equation}
for values of $\beta \geq 1$. If $p_{T|X}$ attains the maximum in $\IB(R)$, then it also attains a minimum of $\mathcal L$ at $\beta = \left(\dv{\IB}{R}(R)\right)^{-1}$ (that is, the reciprocal of the IB curve's slope)\footnote{In general, $\IB(R)$ is not necessarily differentiable, however it is always concave and thus $p_{T|X}$ attains a minimum of $\mathcal L$ at all values of $\beta$, such that $\beta^{-1} \in \partial\IB(R)$, where $\partial\IB(R)$ denotes the superdifferential of the IB function at $R$.} \cite{tishby99ib}.

The IB function is concave in $R$, strictly increasing from $R=0$, at which $\IB(R)=0$, to some $R=R_{\max} \leq H(X)$, and $\IB(R)=I(X;Y)$ for $R\geq R_{\max}$ \cite{witsenhausen1975conditional}. In particular, if $Y$ is a deterministic function of $X$, then $\IB(R)$ is linear for $R\in[0, H(Y)]$ and constant $\IB(R) = H(Y)$ for $R \geq H(Y)$ (see, e.g., \cite{kolchinsky2018caveats}).

Notice that the cardinality (alphabet size) of $T$ is not specified in the optimization problem \eqref{eq:IB}. However, it is essential for solving (or approximating) the maximization in \eqref{eq:IB}. In this work, we are concerned with the minimal alphabet size needed to attain the maximum in $\IB(R)$. 
It was shown in \cite{witsenhausen1975conditional}, as a consequence of an extension of Carath\'eodory's theorem, that an alphabet of size $|\XX| + 1$ suffices for $T$ in order to attain the maximum in $\IB(R)$ for all values of $R$. However, it was unclear that this bound is tight. Note that for $R=0$ the maximum in $\IB(R)$ can be achieved by taking $T$ to be some constant random variable, that is, using an alphabet of size $1$; whereas for $R = H(X)$ the maximum can be achieved with $T=X$, that is, using an alphabet of size $|\XX|$. One could ask, then, whether an alphabet of size $|\XX|$ is enough for all values of $R \in [0, H(X)]$.

In this paper, we show constructively that the answer to this question is negative, that is, the cardinality bound in \cite{witsenhausen1975conditional} is indeed tight. Specifically, we show that if $p_X$ is uniform over an alphabet of size $|\XX| \geq 3$ and $p_{Y|X}$ is a Hamming channel with nontrivial crossover probability, then there exists $R_c\in (0, H(X))$, such that an alphabet of size $|\XX| + 1$ is necessary for achieving the maximum in $\IB(R)$ for all $R \in (0, R_c)$.

Along with our main result, we provide a complete characterization of the IB functions for Hamming channels with uniform input over $|\XX|\geq 3$, as well as their optimal representations. This result is of importance in its own right, as the IB has been fully characterized only for two other families of distributions -- when $X$ and $Y$ are a doubly symmetric binary source (which can be seen as a special case of Hamming channels, where $|\XX|=2$) \cite{witsenhausen1975conditional} and when $X$ and $Y$ are jointly Gaussian \cite{ChechikGIB2005}.

Moreover, we show that in the Hamming case with $|\XX|\geq 3$, the IB function exhibits two interesting phenomena. First, the function contains a linear segment at the origin, even though the joint distribution of $X$ and $Y$ is of full support. Other known instances of joint distributions for which the IB function contains a linear segment have either deterministic $p_{Y|X}$ or decomposable $p_{X,Y}$ \cite{witsenhausen1975conditional}. Clearly, the joint distributions in both cases have incomplete supports. Thus, our result shows that incomplete support is not necessary for the existence of linear segments.  
Second, the minimal necessary cardinality of the representation is not monotonic in $R$. Intuitively, one would expect that as the representation $T$ becomes more compressed, that is, for lower values of $R$, it would require a smaller alphabet size. Nonetheless, we show that in the considered case, compressing the representation beyond a certain point requires, actually, a larger alphabet.

\subsection{Related Work}
It was shown in \cite{harremoes2007information} that a cardinality of $|\XX|$ is sufficient for $T$ to attain the minima of the Lagrangian $\mathcal L[p_{T|X}, \beta]$ for all $\beta \geq 1$. This, in fact, only implies that the cardinality of $|\XX|$ is sufficient for achieving the maximum in $\IB(R)$ when $\IB(R)$ is \emph{strictly} concave in $R$. 
Note that if the IB curve is not strictly concave, then all the values of $R$ that are mapped to a single linear segment\footnote{The existence of a linear segment is guaranteed by the assumption that the IB curve is not strictly concave.} correspond to the same value of $\beta$. Therefore, this cardinality bound only guarantees that at least for one of those values of $R$ there exists a representation $T$ with cardinality $|\XX|$ that attains the maximum in $\IB(R)$. 
This subtle difference between the cardinality bound for the Lagrangian and that of the IB function has been overlooked in  \cite{asoodeh2020bottleneck}, leading erroneously to the conclusion that  
an alphabet of size $|\XX|$ is sufficient for achieving the maximum in $\IB(R)$ for all $R\in[0, H(X)]$.

In \cite{hirche2020alphabet}, the authors consider approximations to the function $\IB(R)$ while imposing bounds on the cardinality of the representation $T$. It should be noted, however, that in this work we are concerned with the necessary alphabet size to achieve the \emph{exact} solutions to the optimization problem in \eqref{eq:IB}.

Finally, the characterization of the IB function for Hamming channels has already been addressed in \cite{dikshtein2021class}. Nevertheless, as the authors relied on the incorrect bound of \cite{asoodeh2020bottleneck}, their characterization is partly inaccurate.

In this paper, we attempt to correct both the cardinality bound of \cite{asoodeh2020bottleneck}, as well as the subsequent analysis of \cite{dikshtein2021class}.

\subsection{Paper Outline}
We start in Section~\ref{sec:Hamming} by defining Hamming channels and referencing two key findings from \cite{witsenhausen1974entropy}. We then connect these findings to the IB in Section~\ref{sec:Main-result} and prove our main result: the cardinality bound on the IB representations, which was first proposed by \cite{witsenhausen1975conditional} to be $|\XX| + 1$, is in fact tight. In Section~\ref{sec:IB-Hamming}, we extend the findings of \cite{witsenhausen1974entropy}, in order to provide a complete characterization of the IB function for Hamming channels for alphabets of size $|\XX| \geq 3$.

\section{Hamming Channels}\label{sec:Hamming}
Let $\XX=\YY$ with $|\XX| = n \geq 2$ and let $\alpha \in [0, \frac{1}{n-1}]$. The $n$-ary Hamming channel with crossover probability $\alpha$, denoted by $\mathcal H_{n,\alpha}$, is defined by the following conditional distribution:
\begin{equation}
    p_{Y|X}(y | x) = \begin{cases}
        1 - (n-1)\alpha, & x=y \\
        \alpha, & x\neq y,
    \end{cases}
\end{equation}
for all $x\in\XX, y\in\YY$. Note that the cases where $\alpha \in \{0,\frac{1}{n},\frac{1}{n-1}\}$ are qualitatively different; we will thus assume that $\alpha\in(0,\frac{1}{n})\cup(\frac{1}{n}, \frac{1}{n-1})$ and consider the extreme cases separately in Section~\ref{subsec:extreme}.

Let $p_{Y|X} = \mathcal H_{n,\alpha}$. In the rest of this section we are concerned with the output entropy of such Hamming channel, that is, the entropy of $Y$, given an input $X\sim p_X$. Since the entropy function is permutation-invariant, we have $H(Y|X=x) = h_n(\alpha)$ for any $x\in\XX$, where we define
\begin{align}
    h_n(q) &= -\big(1 - (n-1)q \big)\log\big(1 - (n-1)q \big) \nonumber\\
           &\quad - (n-1)q\log q,
\end{align}
for $q \in [0, \frac{1}{n-1}]$. We next show that $h_n(q)$ is invertible on $[0, \frac{1}{n}]$. 
\begin{proposition}\label{prop:h-monotonic}
    The function $h_n(q)$ is strictly increasing for $q\in[0, \frac{1}{n}]$ and strictly decreasing for $q\in[\frac{1}{n}, \frac{1}{n-1}]$.
\end{proposition}
\begin{IEEEproof}
    We have $\dv{h_n}{q}(q) = -(n-1)\log\frac{q}{1-(n-1)q}$. For $q\in(0, \frac{1}{n})$, it follows that $\frac{q}{1-(n-1)q}<1$ and so $\dv{h_n}{q}(q)>0$. Similarly, we get $\dv{h_n}{q}(q)<0$ for $q\in(\frac{1}{n}, \frac{1}{n-1})$.
\end{IEEEproof}

In light of this proposition, there exists an inverse function $h_n^{-1}: [0, \log n] \rightarrow [0, \frac{1}{n}]$.

Now, define for $R \in [0, \log n]$,
\begin{equation}\label{eq:phi}
    \varphi(R) = \varphi(R, \mathcal H_{n, \alpha}) =
    \max_{\substack{p_X:\\
    \log n - H(X) \leq R}} \big( \log n - H(Y) \big).
\end{equation}
The following results about $\varphi(R)$ are due to \cite{witsenhausen1974entropy}.\footnote{Our formulation is a slightly modified version of \cite{witsenhausen1974entropy} and enables us to simplify the analysis in the next section. Moreover, although the original claims refer only to $\alpha\leq\frac{1}{n}$, it can be verified that they hold also for $\alpha\leq\frac{1}{n-1}$.}

\begin{lemma}[{[\citenum{witsenhausen1974entropy}, Lemma~7]}]\label{lem:Witsenhausen}
    Let $R\in[0, \log n]$ and let $p_X$ attain the maximum in the definition of $\varphi(R)$. Then, $p_X$ is a permuted version of $\big( 1-(n-1)\beta, \beta, \dots, \beta \big)$, where $\beta = h_n^{-1}(\log n - R)$. 
\end{lemma}

Note that if $p_X$ is as in the statement of Lemma~\ref{lem:Witsenhausen}, then $p_Y$ -- that is, the output distribution of the channel $p_{Y|X} = \mathcal H_{n,\alpha}$, given the input $X\sim p_X$ -- is the corresponding permuted version of $\big( 1-(n-1)\gamma, \gamma, \dots, \gamma \big)$, where $\gamma = \alpha + (1-n\alpha)\beta$. Consequently,
\begin{equation}\label{eq:phi2}
    \varphi(R) = \log n - h_n(\gamma) .
\end{equation}

\begin{remark}\label{rem:phi-monotonic}
    Together with Proposition~\ref{prop:h-monotonic},  
    \eqref{eq:phi2} implies that $\varphi(R)$ is strictly increasing in $R\in[0, \log n]$.
\end{remark}

\begin{theorem}[{[\citenum{witsenhausen1974entropy}, Theorem~6]}]\label{thm:Witsenhausen}
    For $n\geq 3$, the function $\varphi(R, \mathcal H_{n, \alpha})$ is strictly convex in $R$ in a neighborhood of $R=0$. As a consequence, its upper concave envelope is linear in that region.
\end{theorem}

\section{Main Result}\label{sec:Main-result}
\subsection[The IB Function is the Concave Envelope of phi]{The IB Function is the Concave Envelope of $\varphi$}
Let $p_{Y|X} = \mathcal H_{n, \alpha}$ and $p_X = \mathcal U_n$, that is, the uniform distribution over the alphabet $\XX$ of size $n$. Note this implies that also $p_Y = \mathcal U_n$, and so $H(X) = H(Y) = \log n$. Let $p_{T|X}$ define a representation of $X$ over some finite alphabet $\TT$. We have
\begin{align}
    I(X; T) &= H(X) - H(X|T) \nonumber \\
    &= \log n - \sum_{t\in\TT} p_T(t)\,H(X|T=t) \nonumber \\
    &= \sum_{t\in\TT} p_T(t)\,\big( \log n - H(X|T=t) \big) \label{eq:I(X;T)}\\
\intertext{and similarly}
    I(Y; T) &= \sum_{t\in\TT} p_T(t)\,\big( \log n - H(Y|T=t) \big). \label{eq:I(Y;T)}
\end{align}

While the resemblance of $\IB(R)$ \eqref{eq:IB} and $\varphi(R)$ \eqref{eq:phi} becomes apparent, Theorem~\ref{thm:Witsenhausen} implies that those two functions cannot always coincide, since $\IB(R)$ is concave for all $R\in[0, \log n]$. Nonetheless, as our next result shows, there is a strong connection between the two.

\begin{theorem}\label{thm:IB=phibar}
    Let $p_{Y|X} = \mathcal H_{n, \alpha}$ and $p_X = \mathcal U_n$, and denote the upper concave envelope of $\varphi(R)$ by $\conphi(R)$. Then for all $R\in[0, \log n]$ we have
    \begin{equation}
        \IB(R, \mathcal U_n, \mathcal H_{n,\alpha}) = \conphi(R, \mathcal H_{n, \alpha}).
    \end{equation}
\end{theorem}
\begin{IEEEproof}
Fix $R\in[0, \log n]$. We begin by showing that $\conphi(R) \leq \IB(R)$. For that, consider a representation $T$ over the alphabet $\TT=\XX$, defined by an $n$-ary Hamming channel with crossover probability $\beta$, that is, $p_{T|X} = \mathcal H_{n, \beta}$, where $\beta=h_n^{-1}(\log n - R)$. Note that $p_T$ must be uniform, since $p_X$ is uniform. Moreover, for all $x\in\XX, t\in\TT$, we have $p_{X|T}(x|t) = p_{T|X}(t|x)$, so that also $p_{X|T} = \mathcal H_{n, \beta}$. Hence, for all $t\in\TT$ we have $H(X|T=t) = h_n(\beta) = \log n - R$, so from \eqref{eq:I(X;T)} we get $I(X;T) = R$, and thus from the IB definition \eqref{eq:IB} we have
\begin{equation}\label{eq:IBgeqI(Y;T)}
    I(Y; T) \leq \IB(R).
\end{equation}

Now, note that $p_{Y|T} = \mathcal H_{n, \gamma}$, where $\gamma = \alpha + (1-n\alpha)\beta$, so for all $t\in\TT$ we have $H(Y|T=t) = h_n(\gamma)$. Consequently, from \eqref{eq:I(Y;T)} we get $I(Y; T) = \log n - h_n(\gamma)$, and combining this result with \eqref{eq:phi2} and \eqref{eq:IBgeqI(Y;T)} yields
\begin{equation}
    \varphi(R) = \log n - h_n(\gamma) =  I(Y; T) \leq \IB(R).
\end{equation}
Finally, since the IB function is concave, we have $\conphi(R) \leq \IB(R)$.

Conversely, let $p_{T|X}:\XX\rightarrow\TT$ be a maximizer of $\IB(R)$, so that $I(X; T) \leq R$ and $I(Y; T) = \IB(R)$. For each $t\in\TT$ denote $R_t = \log n - H(X|T=t)$, then by \eqref{eq:I(X;T)} we have $\sum_{t\in\TT}p_T(t)\,R_t \leq R$. Moreover, from the definition of $\varphi$ \eqref{eq:phi} we have $\log n - H(Y|T=t) \leq \varphi(R_t)$, so from \eqref{eq:I(Y;T)} we get
\begin{subequations}\label{eq:IBleqphi1}
\begin{align}
    \IB(R) = I(Y; T) &\leq \sum_{t\in\TT}p_T(t)\,\varphi(R_t) \label{eq:leqA}\\
    &\leq \sum_{t\in\TT}p_T(t)\,\conphi(R_t). \label{eq:leqB}
\end{align}
\end{subequations}
Finally, since $\varphi$ is monotonically increasing (see Remark~\ref{rem:phi-monotonic}), so is $\conphi$. Therefore, from the concavity and monotonicity of $\conphi$ we have
\begin{subequations}\label{eq:IBleqphi2}
\begin{align}
    \sum_{t\in\TT}p_T(t)\,\conphi(R_t) &\leq \conphi\left( \sum_{t\in\TT}p_T(t)\,R_t \right) \label{eq:leqC}\\
    &\leq \conphi(R), \label{eq:leqD}
\end{align}
\end{subequations}
which together with \eqref{eq:IBleqphi1} yields $\IB(R) \leq \conphi(R)$, concluding the proof.
\end{IEEEproof}

An immediate consequence of Theorem~\ref{thm:IB=phibar} is that all the inequalities in \eqref{eq:IBleqphi1} and \eqref{eq:IBleqphi2} become equalities, which in turn entail the following implications:
\begin{enumerate}
    \renewcommand{\labelenumi}{(\roman{enumi})}
    \renewcommand{\theenumi}{(\roman{enumi})}
    \item From equalities at \eqref{eq:leqA} and \eqref{eq:leqB} we have $\log n - H(Y|T=t) = \varphi(R_t) = \conphi(R_t)$ for all $t\in\TT$. \label{item:1}
    \item From \ref{item:1} and Lemma~\ref{lem:Witsenhausen}, we have for all $t\in\TT$ that $p_{X|T=t}$ is a permuted version of $\big( 1-(n-1)\beta_t, \beta_t, \dots, \beta_t \big)$, where $\beta_t = h_n^{-1}(\log n - R_t)$. \label{item:2}
    \item From equality at \eqref{eq:leqD}, as $\conphi$ is strictly increasing, we have $I(X; T) = \sum_{t\in\TT}p_T(t)\,R_t = R$. \label{item:3}
    \item From equality at \eqref{eq:leqC} and Jensen's inequality, as $\conphi$ is concave, we have that $R_t = R$ for all $t\in\TT$ or that $\conphi$ is affine (linear) on an interval containing all the $R_t$'s. \label{item:4}
\end{enumerate}

\subsection{Proof of the Cardinality Bound}\label{subsec:cardinality-bound}
Assume that $n\geq 3$ and define
\begin{equation}\label{eq:Rc}
    R_c = \min \{ R > 0 \mid \varphi(R) = \conphi(R) \}.
\end{equation}
Theorem~\ref{thm:Witsenhausen} guarantees that such point $R_c$ exists. Let $R\in(0, R_c)$, and consider a representation $T$ over some finite alphabet $\TT$, such that $p_{T|X}$ is a maximizer of $\IB(R)$.

Note that by the definition of $R_c$, we have $\varphi(R)<\conphi(R)$ for all $R\in(0, R_c)$. Hence, following the notation above, \ref{item:1} implies that for all $t\in\TT$, either $R_t=0$ or $R_t \geq R_c$. Moreover, from \ref{item:3} and the choice of $R\in(0, R_c)$, there must exist $t_0, t_1\in\TT$ such that $R_{t_0} = 0$ and $R_{t_1}\geq R_c>0$. Therefore, $\beta_{t_1} = h_n^{-1}(\log n - R_{t_1})<\frac{1}{n}$ and $\beta_{t_0} = h_n^{-1}(\log n) = \frac{1}{n}$, and so by \ref{item:2}, $p_{X|T}(x|t_0) = \frac{1}{n}$ for all $x\in\XX$.

Now, assume that $|\TT| \leq n$. Since $|\XX| = n$ and $|\TT \setminus \{t_0\}|<n$, \ref{item:2} implies the existence of $x_0\in\XX$, such that $p_{X|T}(x_0|t) = \beta_t\leq\frac{1}{n}$ for all $t\in\TT \setminus \{t_0\}$. Therefore, we get
\begin{align}
    p_X(x_0) &= \sum_{t\in\TT} p_{X|T}(x_0|t)\,p_T(t) \nonumber \\
    &= \frac{1}{n} p_T(t_0) + \beta_{t_1}\,p_T(t_1) + \sum_{\mathclap{t\in\TT \setminus \{t_0, t_1\}} }\beta_t\,p_T(t) \nonumber \\
    &\leq \frac{1}{n} \big(1-p_T(t_1)\big) + \beta_{t_1}\,p_T(t_1) < \frac{1}{n},
\end{align}
because $\beta_{t_1} < \frac{1}{n}$, contradicting the fact that $p_X$ is uniform. We have thus proved by contradiction our main result:
\begin{theorem}\label{thm:main-result}
    Let $|\XX| = n \geq3$ and let $p_{Y|X} = \mathcal H_{n, \alpha}$ and $p_X = \mathcal U_n$. There exists $R_c > 0$, such that for all $R\in(0, R_c)$, if $p_{T|X}:\XX\rightarrow\TT$ is a maximizer of $\IB(R)$, then $|\TT| \geq n + 1$.
\end{theorem}

\section{The Information Bottleneck with Hamming Channels}\label{sec:IB-Hamming}
In this section, we provide a full characterization of the IB function and its corresponding optimal representations, in the case where $|\XX| \geq 3$, $p_X$ is uniform and $p_{Y|X}$ is a Hamming channel. We begin by assuming, as above, that $\alpha\in(0, \frac{1}{n})\cup(\frac{1}{n}, \frac{1}{n-1})$; the extreme cases $\alpha \in \{0, \frac{1}{n}, \frac{1}{n-1}\}$ are treated separately in Section~\ref{subsec:extreme}.

\subsection{The IB Function}
The following lemma extends the findings of Theorem~\ref{thm:Witsenhausen}, showing that $\varphi(R)$ is first strictly convex and then strictly concave.
\begin{lemma}\label{lem:concaveconvex}
    Let $n \geq 3$ and denote $\varphi(R)=\varphi(R, \mathcal H_{n, \alpha})$. Then there exists a unique $R_s\in(0,\log n)$, such that $\dv[2]{\varphi}{R} (R_s)=0$. Moreover, $\dv[2]{\varphi}{R}(R)>0$ for $R\in(0,R_s)$ and $\dv[2]{\varphi}{R}(R)<0$ for $R\in(R_s,\log n)$.
\end{lemma}
The proof is rather technical, and is thus deferred to Appendix~\ref{ap:proof-concaveconvex}. It consists mainly of repeated differentiation and careful consideration of the limits at the boundaries.

Now, with Lemma~\ref{lem:concaveconvex}, we can give an exact description of the upper concave envelope of $\varphi(R)$, that is $\conphi(R)$. Consequently, by Theorem~\ref{thm:IB=phibar}, we get the desired characterization of $\IB(R)$.

\begin{figure}
    \centering
    \includegraphics[width=0.35\textwidth]{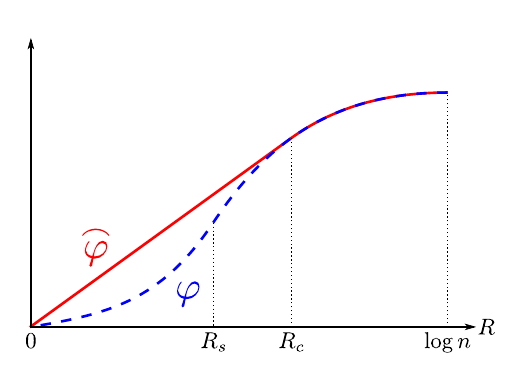}
    \caption{A schematic illustration of $\varphi$ (blue, dashed) and $\conphi$ (red).}
    \label{fig:phibar}
\end{figure}

\begin{figure}
    \centering
    \includegraphics[width=0.45\textwidth]{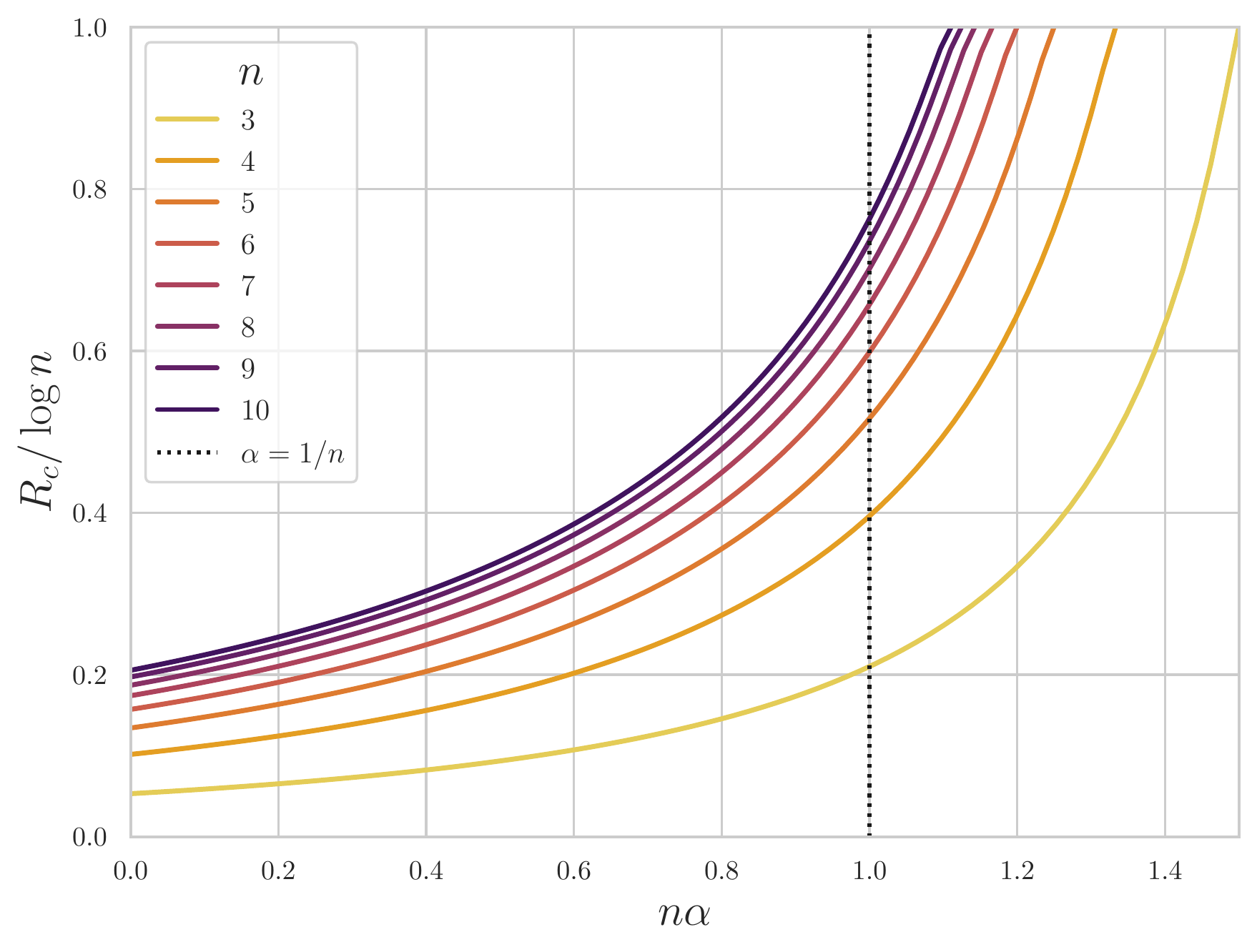}
    \includegraphics[width=0.45\textwidth]{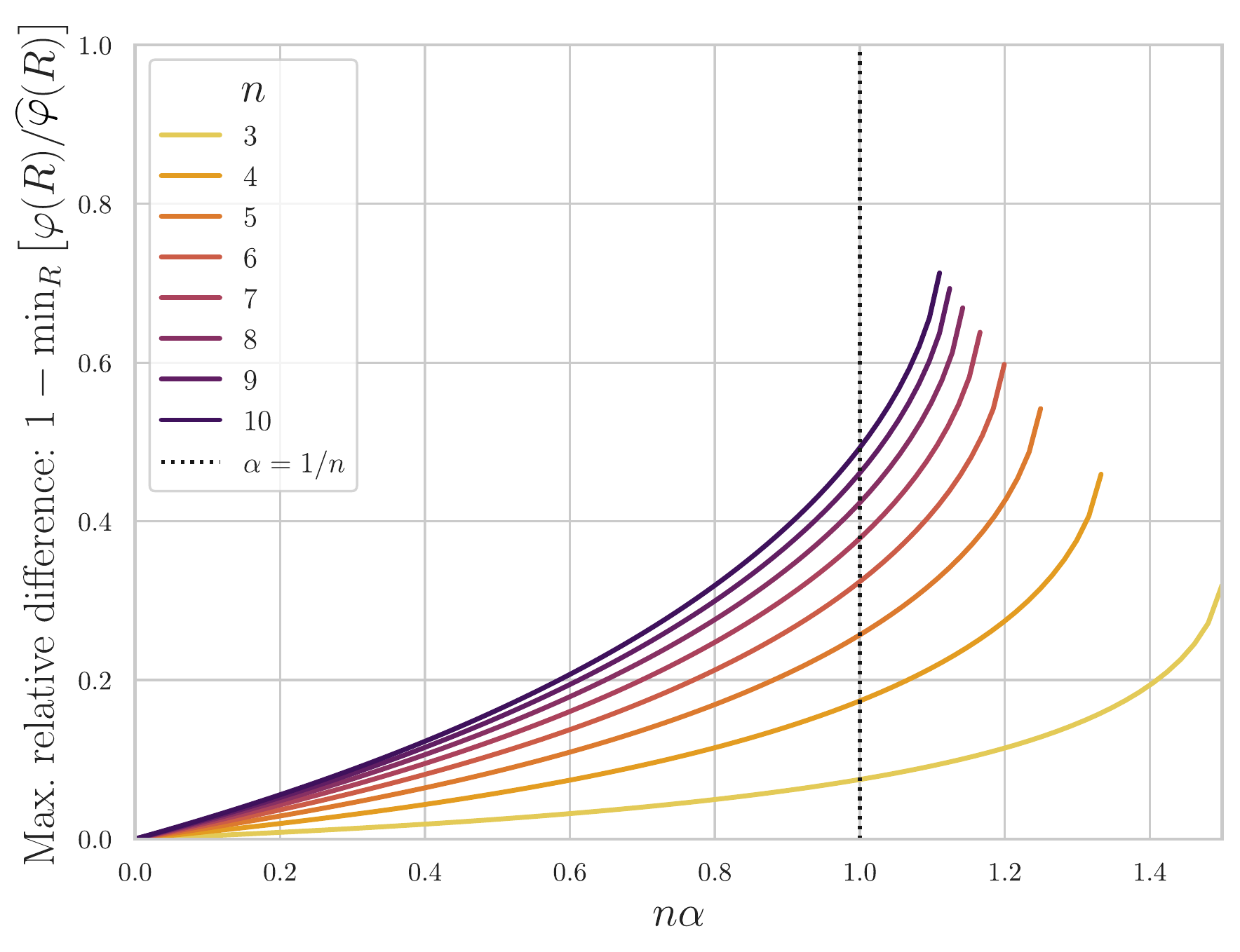}
    \caption{Top: The relative position of $R_c$ in the $R$ axis (that is, the fraction of $\varphi$ that is strictly below $\conphi$ or, equivalently, the fraction of $\conphi$ that is linear; see Fig.~\ref{fig:phibar}). Bottom: The maximal relative difference between $\varphi$ and $\conphi$. Both plots shown at various values of $n$ and $\alpha\in\big(0,\frac{1}{n}\big)\cup\big(\frac{1}{n}, \frac{1}{n-1}\big)$.}
    \label{fig:Rc-maxdiff}
\end{figure}

\begin{theorem}\label{thm:phibar}
    Let $n \geq 3$ and denote by $\conphi(R)$ the concave envelope of $\varphi(R, \mathcal H_{n, \alpha})$. Then we have
    \begin{equation}\label{eq:phibar}
        \conphi(R) = \begin{cases}
            \frac{\varphi(R_c)}{R_c}R, & R\in[0,R_c] \\
            \varphi(R), & R\in(R_c,\log n] ,
        \end{cases}
    \end{equation}
    where $R_c$ is the unique real number in $(0,\log n)$ satisfying $\dv{\varphi}{R}(R_c)=\frac{\varphi(R_c)}{R_c}$. Moreover, $R_c>R_s$.
\end{theorem}
The proof is given in Appendix~\ref{ap:proof-phibar}. Note, in particular, that $\varphi(R) < \conphi(R)$ iff $R\in(0, R_c)$, so $R_c$ coincides with the definition in \eqref{eq:Rc}. A schematic illustration of $\conphi$ is shown in Fig.~\ref{fig:phibar}. In Fig.~\ref{fig:Rc-maxdiff} we show that the difference between $\varphi$ and $\conphi$ can be considerable, especially for large values of $n$ and $\alpha$.

\begin{corollary}
    Let $|\XX| = n \geq3$ and let $p_{Y|X} = \mathcal H_{n, \alpha}$ and $p_X = \mathcal U_n$. Then for all $R\in[0, \log n]$ the function $\IB(R)$ is described by \eqref{eq:phibar}.
\end{corollary}
\begin{IEEEproof}
    Follows immediately from Theorem~\ref{thm:IB=phibar} and Theorem~\ref{thm:phibar}.
\end{IEEEproof}

\subsection{Optimal Representations}
Let $T$ be a representation over a finite alphabet $\TT$, such that $p_{T|X}$ is a maximizer of $\IB(R)$ for some $R\in[0, \log n]$. In what follows, we characterize the representations of minimal cardinality, meaning that there is no representation over a smaller alphabet that maximizes $\IB(R)$.

\begin{theorem}
    Let $|\XX| = n \geq3$ and let $p_{Y|X} = \mathcal H_{n, \alpha}$ and $p_X = \mathcal U_n$. For $R\in[0, \log n]$, let $p_{T|X}:\XX\rightarrow\TT$ be a maximizer of $\IB(R)$ of minimal cardinality, then:
    \begin{enumerate}
    \renewcommand{\labelenumi}{\alph{enumi})}
    \renewcommand{\theenumi}{\alph{enumi})}
        \item If $R=0$, $T$ is achieved by a constant representor, that is, $|\TT| = 1$. \label{item:a}
        \item If $R\in(0, R_c)$, then $|\TT| = n+1$ and $T$ is achieved by time sharing a constant representor $t_0\in\TT$ with probability $1-\frac{R}{R_c}$, and a Hamming channel $\mathcal H_{n,\beta_c}$ with probability $\frac{R}{R_c}$, where $\beta_c=h_n^{-1}(\log n - R_c)$. \label{item:b}
        \item If $R\in[R_c, \log n]$, then $|\TT|=n$ and $T$ is achieved by a Hamming channel $\mathcal H_{n,\beta}$, where $\beta=h_n^{-1}(\log n - R)$. \label{item:c}
    \end{enumerate}
\end{theorem}
\begin{IEEEproof}
\ref{item:a} The case $R=0$ is trivial.

\ref{item:b} Let $R\in(0, R_c)$. By the cardinality bound of \cite{witsenhausen1975conditional}, we have $|\TT| \leq n + 1$. Hence, from Theorem~\ref{thm:main-result}, we get $|\TT| = n + 1$. Now, recall from the analysis in Section~\ref{subsec:cardinality-bound} that for all $t\in\TT$, either $R_t=0$ or $R_t\geq R_c$, and there exist $t_0, t_1\in\TT$ such that $R_{t_0} = 0$ and $R_{t_1}\geq R_c > 0$. Since not all the $R_t$'s are equal, we have from \ref{item:4} that $\conphi$ must be linear on an interval containing $R_t$ for all $t\in\TT$. By Theorem~\ref{thm:phibar}, $\conphi$ is linear only on $[0, R_c]$, and thus we have for all $t\in\TT$, either that $R_t=0$ or $R_t = R_c$.

Note, however, that for all $t\in\TT$ such that $R_t=0$ we have from \ref{item:2} that $\beta_t=\frac{1}{n}$, and thus $p_{X|T}(x|t)=\frac{1}{n}$ for all $x\in\XX$. Therefore, the minimality of $|\TT|$ implies that only $R_{t_0}=0$, since otherwise all the representors with $R_t=0$ could be merged, as their conditional distribution given $X$ is identical. If so, we have $\beta_{t_0} = \frac{1}{n}$, and $\beta_t=\beta_c=h_n^{-1}(\log n - R_c)<\frac{1}{n}$ for all $t\in\TT\setminus\{t_0\}$. Moreover, for all $t\in\TT\setminus\{t_0\}$, it follows from \ref{item:2} that $p_{X|T=t}$ is some permuted version of $\big( 1-(n-1)\beta_c, \beta_c, \dots, \beta_c \big)$. Again, the minimality of $|\TT|$ implies that each permutation occurs not more than once, and since $|\TT\setminus\{t_0\}|=n$ and there are exactly $n$ possible permutations, we get that each of the permuted versions occurs exactly once. In other words, restricted to $\TT\setminus\{t_0\}$, the channel $p_{X|T}$ is an $n$-ary Hamming channel with crossover probability $\beta_c$.

Now, let $t\in\TT\setminus\{t_0\}$, and let $x_0\in\XX$ such that $p_{X|T}(x_0|t) = 1-(n-1)\beta_c$, then
\begin{align}
    p_X(x_0) &= \sum_{t'\in\TT} p_{X|T}(x|t')\,p_T(t') \nonumber \\
    &= \frac{1}{n} p_T(t_0) + \big(1-(n-1)\beta_c\big) p_T(t) + \beta_c \sum_{\mathclap{t'\in\TT \setminus \{t_0, t\}} } p_T(t') \nonumber \\
    &= \frac{1}{n} p_T(t_0) + (1-n\beta_c) p_T(t) + \beta_c \sum_{\mathclap{t'\in\TT \setminus \{t_0\}} } p_T(t') \nonumber \\
    &= \frac{1}{n} p_T(t_0) + (1-n\beta_c) p_T(t) + \beta_c \big(1-p_T(t_0)\big) .
\end{align}
Since $p_X(x_0)=\frac{1}{n}$, we get that
\begin{equation}\label{eq:pT}
    p_T(t) = \frac{1}{n}\big(1-p_T(t_0)\big) .
\end{equation}
Therefore, by Bayes' law, for all $t\in\TT\setminus\{t_0\}$ and $x\in\XX$ we have $p_{T|X}(t|x) = \frac{p_T(t)}{p_X(x)}p_{X|T}(x|t) = \big(1-p_T(t_0)\big) p_{X|T}(x|t)$, meaning that, conditioned on $\TT\setminus\{t_0\}$, the channel $p_{T|X}$ is also an $n$-ary Hamming channel with crossover probability $\beta_c$.

To conclude this part of the proof, we need to show that $p_T(t_0) = 1-\frac{R}{R_c}$. Indeed, by \ref{item:3}, we have
\begin{align}
    R &= \sum_{t\in\TT} p_T(t)\,R_t \nonumber \\
    &= p_T(t_0)\,R_{t_0} + R_c\sum_{\mathclap{t\in\TT \setminus \{t_0\}} } p_T(t) \nonumber \\
    &= R_c\big(1-p_T(t_0)\big),
\end{align}
so $p_T(t_0) = 1-\frac{R}{R_c}$, and thus by \eqref{eq:pT} we get also that $p_T(t) = \frac{R}{nR_c}$ for all $t\in\TT\setminus\{t_0\}$.

\ref{item:c} Assume that $R\in[R_c, \log n]$ and note that from \ref{item:3} we have $R=\sum_{t\in\TT} p_T(t)\,R_t$. If $R_t\leq R_c$ for all $t\in\TT$, then it must hold that $R_t=R_c=R$ for all $t\in\TT$; otherwise we get $R<R_c$, contradicting the assumption. If, on the other hand, there exist $t\in\TT$ such that $R_t > R_c$, then from \ref{item:4} we get $R_t=R$ for all $t\in\TT$, since by Theorem~\ref{thm:phibar} and Lemma~\ref{lem:concaveconvex}, $\conphi$ is strictly concave around $R_t$. Therefore, in any case, $R_t=R > 0$ for all $t\in\TT$.

Now, from \ref{item:2}, it follows that $p_{X|T=t}$ is a permuted version of $\big( 1-(n-1)\beta, \beta, \dots, \beta \big)$ for all $t\in\TT$, where $\beta = h_n^{-1}(\log n - R)<\frac{1}{n}$. As above, the minimality of $|\TT|$ implies that each permutation occurs not more than once, hence $|\TT| \leq n$. However, if $|\TT|<n$ then there exists $x_0\in\XX$, such that $p_{X|T}(x_0|t) = \beta$ for all $t\in\TT$, and thus $p_X(x_0) = \sum_{t\in\TT}p_{X|T}(x_0|t)\,p_T(t) = \beta <\frac{1}{n}$, contradicting the fact that $p_X$ is uniform. Consequently, $|\TT| = n$ and each of the permuted versions occurs exactly once. In other words, the channel $p_{X|T}$ is an $n$-ary Hamming channel with crossover probability $\beta$.

Finally, let $t\in\TT$, and let $x_0\in\XX$ such that $p_{X|T}(x_0|t) = 1-(n-1)\beta$, then
\begin{align}
    p_X(x_0) &= \sum_{t'\in\TT} p_{X|T}(x|t')\,p_T(t') \nonumber \\
    &= \big(1-(n-1)\beta\big) p_T(t) + \beta \sum_{\mathclap{t'\in\TT \setminus \{t\}} } p_T(t') \nonumber \\
    &= (1-n\beta) p_T(t) + \beta \sum_{\mathclap{t'\in\TT} } p_T(t') \nonumber \\
    &= (1-n\beta) p_T(t) + \beta .
\end{align}
Since $p_X(x_0)=\frac{1}{n}$, we get that $p_T(t)=\frac{1}{n}$. Therefore, by Bayes' law, for all $t\in\TT$ and $x\in\XX$ we have $p_{T|X}(t|x) = \frac{p_T(t)}{p_X(x)}p_{X|T}(x|t) = p_{X|T}(x|t)$, meaning that $p_{T|X} = \mathcal H_{n,\beta}$, where $\beta=h_n^{-1}(\log n - R)$.
\end{IEEEproof}

\subsection[Extreme Cases of alpha]{Extreme Cases of $\alpha$}\label{subsec:extreme}
Here we consider the three extreme cases $\alpha \in \{0, \frac{1}{n}, \frac{1}{n-1}\}$. When $\alpha=0$, the channel $P_{Y|X}$ is deterministic and we have $\IB(R)=R$ for all $R\in[0, \log n]$ (see, e.g., \cite{kolchinsky2018caveats}). When $\alpha=\frac{1}{n}$, we have $p_{Y|X}(y|x)=\frac{1}{n}=p_Y(y)$ for all $x\in\XX,y\in\YY$, meaning that $X$ and $Y$ are independent, and thus by the data processing inequality $0\leq I(T;Y)\leq I(X;Y)=0$ for all Markov chains $Y-X-T$. As a consequence, $\IB(R)=0$ for all $R\in[0, \log n]$.

Finally, when $\alpha=\frac{1}{n-1}$, we have $\dv[2]{\varphi}{R}(R) < 0$ for all $R\in(0, \log n)$. Consequently, $\IB(R) = \conphi(R) = \frac{\log \frac{n}{n-1}}{\log n} R$ for all $R\in[0, \log n]$. Interestingly, the entire IB curve in this case is linear, just like in the deterministic case ($\alpha=0$).

%%%%%%
%% To balance the columns at the last page of the paper use this
%% command somewhere at the top of the first column of the last page:
%%
% \enlargethispage{-5cm} 
%%
%% where the exact amount of page reduction has to be adapted to the
%% actual situation.
%%
%% If the balancing should occur in the middle of the references, use
%% the following trigger:
%%
% \IEEEtriggeratref{3}
%%
%% which triggers a \newpage (i.e., new column) just before the given
%% reference number. Note that you need to adapt this if you modify
%% the paper. The "triggered" command can be changed if desired:
%%
% \IEEEtriggercmd{\enlargethispage{-20cm}}
%%
%%%%%%

%%%%%%
%% References:
%% We recommend the usage of BibTeX:
%%
%\bibliographystyle{IEEEtran}
%\bibliography{definitions,bibliofile}
%%
%% where we here have assume the existence of the files
%% definitions.bib and bibliofile.bib.
%% BibTeX documentation can be obtained at:
%% http://www.ctan.org/tex-archive/biblio/bibtex/contrib/doc/
%%%%%%
%% Or you use manual references (pay attention to consistency and the
%% formatting style!):

\bibliographystyle{IEEEtran}
\bibliography{references}

\appendices

\section{Proof of Lemma~\ref{lem:concaveconvex}}\label{ap:proof-concaveconvex}
\begin{IEEEproof}
Let $n\geq3$ and $\alpha \in (0, \frac{1}{n})$. For $R\in[0, \log n]$, recall that
\begin{equation}
    \varphi(R) = \log n - h_n(\gamma) ,
\end{equation}
where
\begin{align}
    h_n(q) &= -\big(1 - (n-1)q \big)\log\big(1 - (n-1)q \big) \nonumber\\
           &\quad - (n-1)q\log q ,
\end{align}
$\gamma = \alpha + (1-n\alpha)\beta$ and $\beta = h_n^{-1}(\log n - R)$. In the interest of readability, we omit in our notation the explicit dependence of $\beta$ on $R$ and of $\gamma$ on $\beta$ (and thus also on $R$). We have
\begin{align}
    \dv{h_n}{q} &= -(n-1)\log\frac{q}{1-(n-1)q} , \label{eq:dh/dq} \\
    \dv{\beta}{R} &= -\left( \dv{h_n}{q}(\beta) \right)^{-1} = \frac{1}{(n-1)\log\frac{\beta}{1-(n-1)\beta}} , \\
    \dv{\gamma}{\beta} &= 1 - n\alpha , \\
    \dv{\varphi}{\gamma} &= -\dv{h_n}{q}(\gamma) = -(n-1)\log\frac{\gamma}{1-(n-1)\gamma} .
\end{align}
Now, using the chain rule $ \dv{\varphi}{R} = \dv{\varphi}{\gamma}\dv{\gamma}{\beta}\dv{\beta}{R}$, we get
\begin{align}
    \dv{\varphi}{R}(R) &= \frac{(1-n\alpha)\log\frac{\gamma}{1-(n-1)\gamma}}{\log\frac{\beta}{1-(n-1)\beta}} , \label{eq:dphi/dR} \\
\intertext{and then}
    \dv[2]{\varphi}{R}(R) &= \left(\frac{(1-n\alpha)^2}{\gamma(1-(n-1)\gamma)}\log\frac{\beta}{1-(n-1)\beta} \right. \nonumber \\
    &\quad \left. -\frac{1-n\alpha}{\beta(1-(n-1)\beta)}\log\frac{\gamma}{1-(n-1)\gamma} \right) \nonumber \\
    &\quad \left(\log\frac{\beta}{1-(n-1)\beta}\right)^{-3} (n-1)^{-1} .
\end{align}

Note that that for $R\in(0,\log n)$ we have $\beta\in(0,\frac{1}{n})$, so $\frac{\beta}{1-(n-1)\beta}<1$ and thus the sign of $\dv[2]{\varphi}{R}(R)$ is the same as that of
\begin{align}
    \eta(\beta) &= (1-n\alpha)\gamma(1-(n-1)\gamma)\log\frac{\gamma}{1-(n-1)\gamma} \nonumber \\
    &\quad -(1-n\alpha)^2\beta(1-(n-1)\beta)\log\frac{\beta}{1-(n-1)\beta} .
\end{align}
It can be verified that
\begin{align}
    \dv{\eta}{\beta}(\beta) &= (1-n\alpha)^2(1-2(n-1)\gamma)\log\frac{\gamma}{1-(n-1)\gamma} \nonumber \\
    &\quad -(1-n\alpha)^2(1-2(n-1)\beta)\log\frac{\beta}{1-(n-1)\beta} , \\
    \dv[2]{\eta}{\beta}(\beta) &= 2(n-1)(1-n\alpha)^2\log\frac{\beta}{1-(n-1)\beta} \nonumber \\
    &\quad -(1-n\alpha)^2\frac{1-2(n-1)\beta}{\beta(1-(n-1)\beta)} \nonumber \\
    &\quad -2(n-1)(1-n\alpha)^3\log\frac{\gamma}{1-(n-1)\gamma} \nonumber \\
    &\quad +(1-n\alpha)^3\frac{1-2(n-1)\gamma}{\gamma(1-(n-1)\gamma)} , \\
    \dv[3]{\eta}{\beta}(\beta) &= \frac{(1-n\alpha)^2}{\beta^2(1-(n-1)\beta)^2}-\frac{(1-n\alpha)^4}{\gamma^2(1-(n-1)\gamma)^2} .
\end{align}

We shall show that $\dv[3]{\eta}{\beta}(\beta) > 0$ for $\beta \in (0,\frac{1}{n})$. Since
\begin{multline}
    \gamma^2(1-(n-1)\gamma)^2-(1-n\alpha)^2\beta^2(1-(n-1)\beta)^2 \\
    =\big(\gamma(1-(n-1)\gamma)+(1-n\alpha)\beta(1-(n-1)\beta)\big)\\
    \big(\gamma(1-(n-1)\gamma)-(1-n\alpha)\beta(1-(n-1)\beta)\big) ,
\end{multline}
it suffices to show that
\begin{equation}
   \gamma(1-(n-1)\gamma)-(1-n\alpha)\beta(1-(n-1)\beta)>0
\end{equation}
for $\beta\in(0,\frac{1}{n})$, which is indeed true as
\begin{multline}
  \gamma(1-(n-1)\gamma)-(1-n\alpha)\beta(1-(n-1)\beta)\\
  =(\alpha+(1-n\alpha)\beta)(1-(n-1)\alpha-(n-1)(1-n\alpha)\beta)\\
  -(1-n\alpha)\beta(1-(n-1)\beta)\\
  =(n-1)(1-n\alpha)n\alpha\beta^2-2(n-1)(1-n\alpha)\alpha\beta-(n-1)\alpha^2+\alpha\\
  =(n-1)(1-n\alpha)n\alpha(\beta^2-2\frac{\beta}{n}+\frac{1}{n^2})\\
  -\frac{(n-1)(1-n\alpha)\alpha}{n}-(n-1)\alpha^2+\alpha\\
  =(n-1)(1-n\alpha)n\alpha(\beta-\frac{1}{n})^2+\frac{\alpha}{n}\\
  >(n-1)(1-\frac{n}{n-1})n\alpha(\beta-\frac{1}{n})^2+\frac{\alpha}{n}\\
  =-n\alpha(\beta-\frac{1}{n})^2+\frac{\alpha}{n}\\
  >-n\alpha\frac{1}{n^2}+\frac{\alpha}{n} = 0,
\end{multline}
where the first inequality follows from the fact that $\alpha < \frac{1}{n-1}$ and the second from $\beta\in(0,\frac{1}{n})$. Consequently, $\dv[2]{\eta}{\beta}(\beta)$ is strictly increasing in $\beta$.

Now, note that $\lim_{\beta\downarrow 0}\dv[2]{\eta}{\beta}(\beta)=-\infty$ and $\dv[2]{\eta}{\beta}(\frac{1}{n})=(n-2)n^2\alpha(1-n\alpha)^2>0$, as $n\geq 3$. Therefore, $\dv{\eta}{\beta}(\beta)$ is first strictly decreasing in $\beta$ and then strictly increasing in $\beta$. Moreover, we have $\lim_{\beta\downarrow 0}\dv{\eta}{\beta}(\beta)=\infty$ and $\dv{\eta}{\beta}(\frac{1}{n})=0$. This further implies that $\eta(\beta)$ is first strictly increasing in $\beta$ and then strictly decreasing in $\beta$. Since $\lim_{\beta\downarrow 0}\eta(\beta)=(1-n\alpha)\alpha(1-(n-1)\alpha)\log\frac{\alpha}{1-(n-1)\alpha}<0$ and $\eta(\frac{1}{n})=0$, there exists a unique $\beta_s\in(0,\frac{1}{n})$ such that $\eta(\beta_s)=0$. In addition, $\eta(\beta)<0$ for $\beta\in(0,\beta_s)$ and $\eta(\beta)>0$ for $\beta\in(\beta_s,\log n)$. The proof is complete in view of the fact that $R$ is a strictly decreasing function of $\beta$.

Interestingly, for $n=2$ it follows from the analysis above that $\eta(\beta) < 0$ for all $\beta\in(0, \frac{1}{n})$, or equivalently, that $\dv[2]{\varphi}{R}(R) < 0$ for all $R\in(0, \log n)$. This proves that the requirement that $n\geq 3$ is indeed necessary.

\end{IEEEproof}

\section{Proof of Theorem~\ref{thm:phibar}}\label{ap:proof-phibar}
\begin{IEEEproof}
First, note that when $R=0$ we have $\beta = h_n^{-1}(\log n) = \frac{1}{n}$, so $\gamma = \frac{1}{n}$ and we get $\varphi(0)=0$. Since $\varphi$ is strictly convex for $R\in(0,R_s)$, it follows that
\begin{equation}
    \dv{\varphi}{R}(R)>\frac{\varphi(R)}{R}
\end{equation}
for $R\in(0,R_s]$. On the other hand, when $R=\log n$ we have $\beta = 0$, so $\gamma = \alpha < \frac{1}{n}$ and we get $\varphi(\log n) > 0$. In addition, from \eqref{eq:dphi/dR} we have $\dv{\varphi}{R}(\log n)=0$, so together we get
\begin{equation}
    \dv{\varphi}{R}(\log n)<\frac{\varphi(\log n)}{\log n} ,
\end{equation}
which implies the existence of $R_c\in(R_s,\log n)$ with
\begin{equation}
    \dv{\varphi}{R}(R_c)=\frac{\varphi(R_c)}{R_c} .
\end{equation}

To show that such $R_c$ is unique, it suffices to prove that $R\dv{\varphi}{R}(R)-\varphi(R)$ is strictly decreasing for $R\in(R_s,\log n)$, which is indeed true because 
\begin{equation}
    \dv{}{R}(R\dv{\varphi}{R})(R)-\dv{\varphi}{R}(R)=R\dv[2]{\varphi}{R}(R) < 0
\end{equation}
for all $R\in(R_s,\log n)$.

Note that $\varphi(R)=\frac{\varphi(R_c)}{R_c}R$ for $R=0$ and $R=R_c$. As $\dv{\varphi}{R}(R_c)=\frac{\varphi(R_c)}{R_c}$, $\dv[2]{\varphi}{R}(R)>0$ for $R\in(0,R_s)$, and $\dv[2]{\varphi}{R}(R)<0$ for $R\in(R_s,\log n)$, we must have
\begin{equation}
    \varphi(R)<\frac{\varphi(R_c)}{R_c}R
\end{equation}
for $R\in(0,R_c)\cup(R_c,\log n]$. Finally, in view of the fact that $\varphi(R)$ is concave in $R$ for $R\in[R_c,\log n]$, the expression of $\conphi(R)$ is indeed as stated in \eqref{eq:phibar}.
\end{IEEEproof}

\end{document}